\input{aipcheck}

\documentclass[
    ,final            
  ]
  {aipproc}

\layoutstyle{8x11single}

\usepackage[cp1251,engukr]{statphys09}
\begin{document}

\title{Field Theoretical Approach to Bicritical and Tetracritical Behavior: Static and Dynamics}


\author{R. Folk}{
  address={Institute for Theoretical Physics, Johannes Kepler
University Linz, Altenbergerstrasse 69, A-4040, Linz, Austria} }

\author{Yu. Holovatch}{
  address={Institute for Condensed Matter Physics, National Academy
of Sciences of Ukraine, 1~Svientsitskii Str., UA--79011 Lviv,
Ukraine} , altaddress={Institute for Theoretical Physics, Johannes
Kepler University Linz, Altenbergerstrasse 69, A-4040, Linz,
Austria}}

\author{G.Moser}{
  address={Department for Material Research and Physics, Paris Lodron University
Salzburg, Hellbrunnerstrasse 34, A-5020 Salzburg, Austria} }

\begin{abstract}
We discuss the static and dynamic multicritical behavior of
three-dimensional systems of $O(n_\|)\oplus O(n_\perp)$ symmetry as
it is explained by the field theoretical renormalization group
method. Whereas the static renormalization group functions are
currently know within high order expansions, we show that an account
of two loop contributions refined by an appropriate resummation
technique gives an accurate quantitative description of the
multicritical behavior. One of the essential features of the static
multicritical behavior obtained already in two loop order for the
interesting case of an antiferromagnet in a magnetic field
($n_\|=1$, $n_\perp=2$) are the stability of the biconical fixed
point  and the neighborhood of the stability border lines to the
other fixed points leading to very small transient exponents. We
further pursue an analysis of dynamical multicritical behavior
choosing different forms of critical dynamics and calculating
asymptotic and effective dynamical exponents  within the minimal
subtraction scheme.
\end{abstract}

\maketitle


\section{Introduction}

Beneath the milestone contributions of N.N. Bogolyubov that shaped
modern theoretical physics one definitely should mention his and
D.N. Shirkov work on the renormalization group
(RG)\cite{Bogolyubov59}. Three papers on RG written in the
mid-50-ies by three different groups \cite{RGpapers} addressed
quantum electrodynamics problems, however very soon their importance
has been realized in - on the first sight - very different field of
phase transitions and critical phenomena. It is generally recognized
by now that the success in conceptual understanding and quantitative
description of behavior in the vicinity of critical points in
different condensed matter systems is due to the effective
application of the RG ideas originating from the above papers
\cite{note1}. It is our pleasure to contribute to these
Proceedings\footnote{The paper is based on the invited lecture given
by one of us (R.F.) at the Conference Statphys'09 dedicated to the
100-th anniversary of N.N.Bogolyubov (23.06-25.06.2009, Lviv,
Ukraine)} by a short review of recent work done by application of
the field theoretical RG approach to analysis of multicritical
phenomena.

\begin{figure}
\tabcolsep=.05\textwidth
\begin{tabular}{cc}
\includegraphics[width=.4\textwidth]{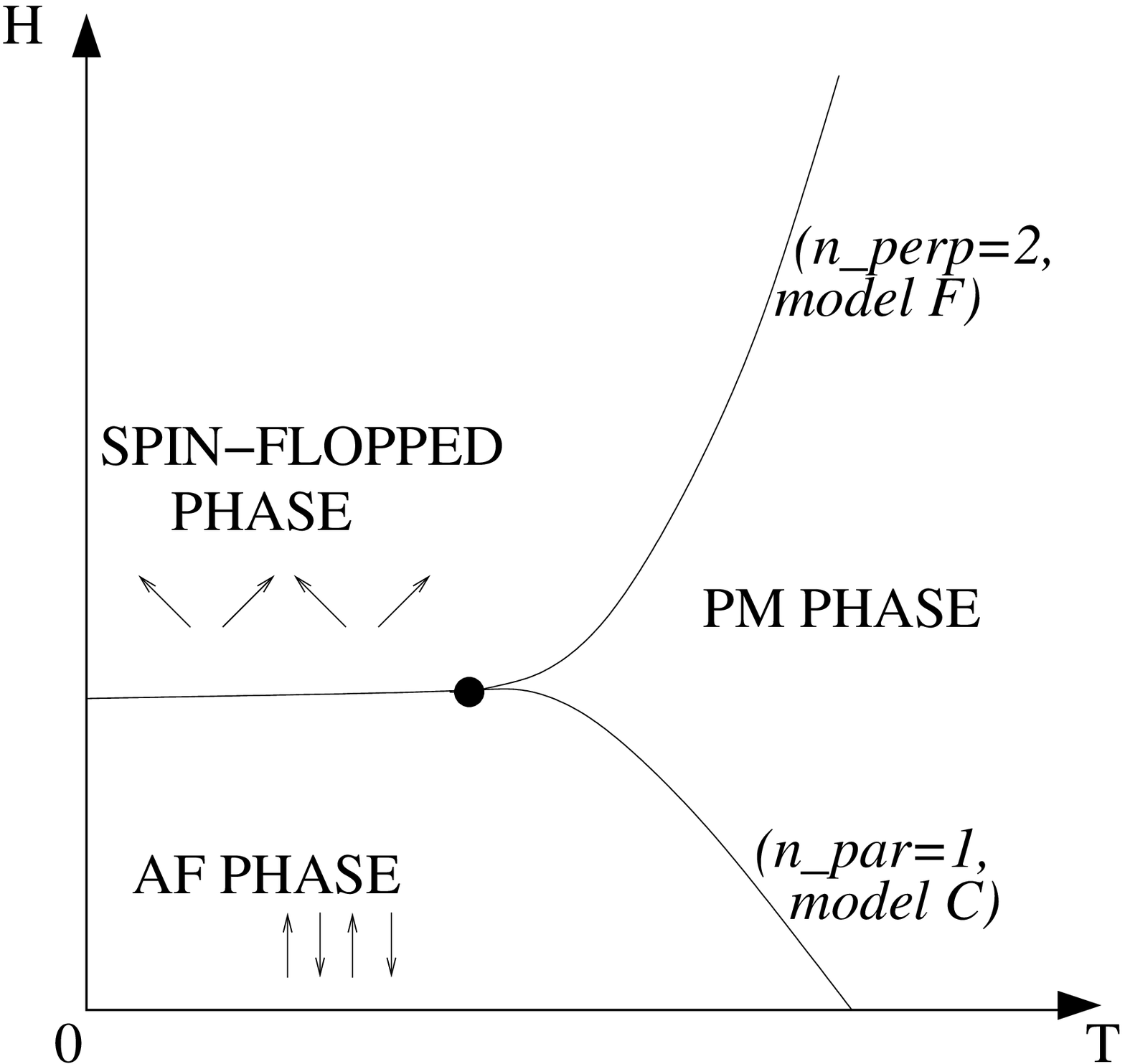} &
\includegraphics[width=.4\textwidth]{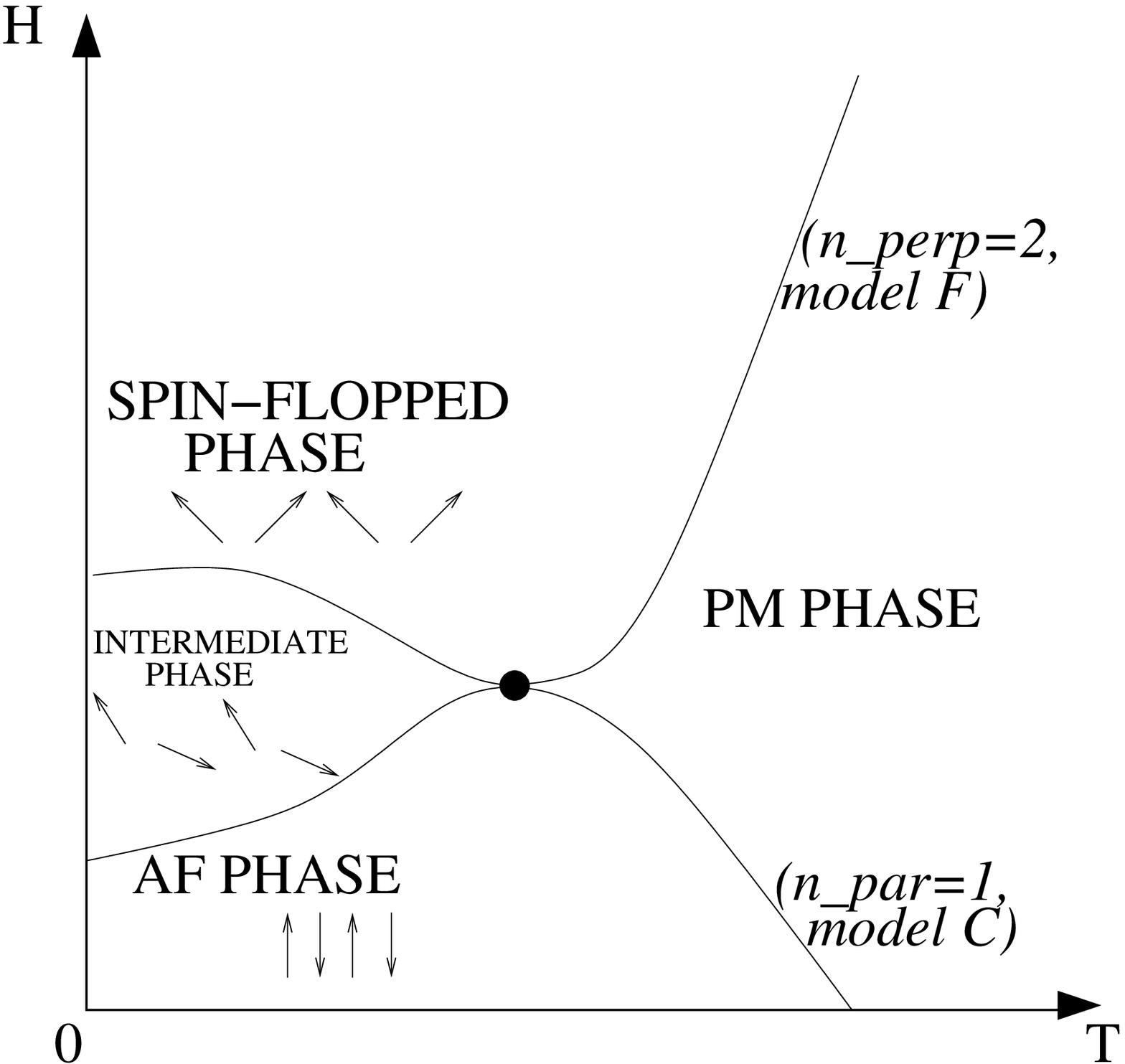} \\
{\bf a.} &  {\bf b.} \\
\end{tabular}
 \caption{\label{fig1} Typical phase diagrams of anisotropic
antiferromagnets   in a uniform parallel external magnetic field
$H$. Types of ordering are schematically shown by arrows.  {\bf a}:
the {\it bicritical} point. Three phases - an antiferromagnetic
phase, a spin flop phase and the paramagnetic phase are in
coexistence. The phase transition lines to the paramagnetic phase
are second order transition lines, whereas the transition line
between the spin flop and the antiferromagnetic phase is of first
order. {\bf b}:  the {\it tetracritical} point. Four phases - an
antiferromagnetic phase, a spin flop phase, an intermediate or mixed
phase and the paramagnetic phase - are in coexistence. All
transition lines are of the second order in this case. Also
indicated is the dynamical universality class of the transition from
the paramagnetic to the corresponding ordered phase according to the
classification of Hohenberg  and Halperin \cite{HHreview} for the
three component antiferromagnet.}
\end{figure}

Multicritical points appear on phase diagrams of various systems
that contain several phase transitions lines. In the vicinity of the
meeting points of such lines the multicritical behavior is observed,
which is characterized by competition of different types of
ordering. Prominent examples are given by the antiferromagnets in an
external magnetic field  like GdALO$_3$,  MnF$_2$, MnCl$_2$4D$_2$O,
Mn$_2$AS$_4$ (A=Si or Ge) \cite{antiferromagnets}. Other examples
are given by the layered cuprate antiferromagnets like
(Ca,La)$_{14}$Cu$_{24}$O$_{41}$. Schematic phase diagrams of such
systems are shown in Fig.\ref{fig1} in a $H$-$T$ plane. There,
multicritical points of two different types are manifested. At a
{\it bicritical} point (Fig. \ref{fig1}{\bf a}) three phases are in
coexistence, whereas four phases coexist in the {\it tetracritical}
point (Fig. \ref{fig1}{\bf b}). On a more general level, the
multicritical behavior is inherent to a critical system when some
"nonordering" field is applied. Such a field (beside the magnetic
field $H$ this may be pressure, stress etc.) may alter non-universal
parameters of the system and lead to appearance of the lines of
phase transition points. Besides the above example that concern the
shift of the N\'eel point of anisotropic antiferromagnets by a
uniform magnetic filed, other examples of multicritical behavior are
observed at a shift of the Curie points under applied pressure or
depression of the $\lambda$ point in ${\rm ^4He}$ at dilution by
${\rm ^3He}$ \cite{Liu72}.


A field theoretic description of multicritical behavior starts with
a static effective Hamiltonian for an $n$-component field
$\Phi=(\vec{\phi}_\|,\vec{\phi}_\perp)$ of $O(n_\|)\oplus
O(n_\perp)$ symmetry ($n_\| + n_\perp=n$).  An account of the
interaction between the two order parameters $\vec{\phi}_\|$ and
$\vec{\phi}_\perp$ leads to different types of multicritical
behavior connected with the stable fixed point (FP) found in the RG
treatment
\cite{Nelson74,Aharony74,Lyuksyutov75,Kosterlitz76,Prudnikov98,Calabrese03,Folk08a}.
In particular, the bicritical point (Fig. \ref{fig1}{\bf a}) has
been connected with the stability of the {\it isotropic Heisenberg}
fixed point of $O(n_\|+n_\perp)$ symmetry, whereas the tetracritical
point (Fig. \ref{fig1}{\bf b}) corresponds to a FP of $O(n_\|)\oplus
O(n_\perp)$ symmetry, which might be either the so called {\it
biconical} FP or the {\it decoupling} FP. In the last FP the
parallel and the perpendicular components of the order parameter are
asymptotically decoupled. If no FP is reached the multicritical
point might be of first order, i.e. a triple point.

Quite recently the possible types of  phase diagrams in the $H-T$
plane of three dimensional  uniaxial anisotropic antiferromagnets
have been studied by Monte Carlo simulations \cite{Selke}. For
$n_\|=1$ and $n_\perp=2$ a phase diagram with a bicritical point has
been found in agreement with earlier simulations \cite{Landau78},
but contrary to the results of RG theory in higher loop orders
\cite{Calabrese03}.

The dynamics of antiferromagnets in a magnetic field is quite
complicated.  To account for conservation laws present in such
systems, the dynamical equations of motion should contain  coupling
terms between the two order parameters (the components of the
staggered magnetization parallel and perpendicular to the magnetic
field, $\vec{\phi}_\|$ and $\vec{\phi}_\perp$) and conserved
densities (e.g. the parallel component of the magnetization or the
energy density).  First formulation of the equations of motion at
multicritical points has been done in Ref. \cite{Dohm77}. The
simplest form of dynamics assumes the relaxational behavior for the
two order parameters $\vec{\phi}_\|$ and $\vec{\phi}_\perp$ (the
so-called model A) \cite{HHreview,Halperin74}. Dynamical
multicritical behavior within the one-loop approximation has been
considered in \cite{Dohm77} on the basis of the static one loop
results \cite{Kosterlitz76}. A further step to the complete model is
to include the diffusive dynamics of the slow conserved density
leading to a model C like extension. This model has been studied in
one loop order in Refs. \cite{Dohm77,Dohm79,Dohm83} taking into
account only a part of dynamical two loop order terms and one loop
statics. In order to get more insight in the dynamics in the
vicinity of multicritical points, recently we have reconsidered the
above dynamical models within the two loop approximation
\cite{Folk08b,Folk09a}.

In what follows below we briefly summarize an outcome of an RG
analysis of multicritical behavior paying special attention to an
impact of the non-universal contributions to an asymptotic behavior.
In particular, we will show that an account of two loop part of the
RG expansions refined by an appropriate resummation technique gives
an accurate quantitative description of the static multicritical
behavior. Furthermore, we pursue an analysis of dynamical
multicritical behavior choosing different forms of critical dynamics
and calculating asymptotic and effective dynamical exponents.

\section{RG flows and static multicritical behavior}

The generalized static $O(n_\|)\oplus O(n_\perp)$-symmetrical
effective Hamiltonian  that results from the decomposition of the
$n$-component order parameter field into two mutually interacting
fields $\vec{\phi}_\|$ and $\vec{\phi}_\perp$ of different
irreducible representations of dimensions $n_\|$ and $n_\perp$,
$n=n_\|+n_\perp$, reads:
\begin{eqnarray}\label{1}
{\cal H}\!&=&\!\int\!
d^dx\Bigg\{\frac{1}{2}\mathring{r}_\perp\vec{\phi}_{\perp 0}
\cdot\vec{\phi}_{\perp
0}+\frac{1}{2}\sum_{i=1}^{n_\perp}\nabla_i\vec{\phi}_{\perp 0}\cdot
\nabla_i\vec{\phi}_{\perp 0}
+\frac{1}{2}\mathring{r}_\|\vec{\phi}_{\| 0} \cdot\vec{\phi}_{\|
0}+\frac{1}{2}\sum_{i=1}^{n_\|}\nabla_i\vec{\phi}_{\| 0}\cdot
\nabla_i\vec{\phi}_{\| 0}  \nonumber \\
&+&\frac{\mathring{u}_\perp}{4!}\Big(\vec{\phi}_{\perp
0}\cdot\vec{\phi}_{\perp 0}\Big)^2
+\frac{\mathring{u}_\|}{4!}\Big(\vec{\phi}_{\| 0}\cdot\vec{\phi}_{\|
0}\Big)^2
+\frac{2\mathring{u}_\times}{4!}\Big(\vec{\phi}_{\perp
0}\cdot\vec{\phi}_{\perp 0}\Big) \Big(\vec{\phi}_{\|
0}\cdot\vec{\phi}_{\| 0}\Big) \Bigg\} \ .
\end{eqnarray}
Here,
$\{\mathring{u}_\perp,\mathring{u}_\times,\mathring{u}_\|\}=\{\mathring{u}\}$
and $\mathring{r}_\perp$, $\mathring{r}_\|$ are couplings and
masses, correspondingly, index 0 refers to the bare quantities, and
central dots stand for scalar products. The decomposition in
parallel and perpendicular order parameter components allows to
describe the multicritical behavior at the meeting point of two
critical lines: (i) the line where $\mathring{r}_\perp$ becomes zero
and the $n_\perp$-dimensional components $\vec{\phi}_{\perp 0}$ are
the order parameter, and (ii) the line where $\mathring{r}_\|$
becomes zero and the order parameter is $\vec{\phi}_{\|0}$. At the
meeting point both quadratic terms become zero and both components
of $\vec{\phi}_{0}$ have to be taken into account. As has been
predicted already by the one-loop RG analysis
\cite{Lyuksyutov75,Kosterlitz76}, an effective Hamiltonian (\ref{1})
describes three different types of multicritical behavior that are
governed by three different FPs: (i) the isotropic $n$ component
Heisenberg FP, called below ${\cal H}(n)$, all fourth order
couplings are equal in this FP, (ii) the decoupling FP point ${\cal
D}$, which consists of a combination of the FPs ${\cal H}(n_\perp)$
and ${\cal H}(n_\|)$ of two decoupled systems and (iii) the
biconical FP, ${\cal B}$, with nontrivial nonzero couplings. As it
was revealed by subsequent  calculations
\cite{Prudnikov98,Calabrese03} the FP picture does not change
qualitatively with an account of higher orders of the perturbation
theory. However, the one-loop results attain essential quantitative
changes that lead to drastic modification of the type of a phase
diagram. Typical example may be given by the behavior of the
$\beta$-functions, that describe  flow of the fourth order couplings
$\{\mathring{u}\}$ under renormalization. The above functions,
calculated in the two-loop approximation with the minimal
subtraction RG scheme read \cite{Folk08a}:
\begin{eqnarray}
\label{2} \beta_{u_\perp}&=& -\varepsilon  u_\perp +
\frac{(n_\perp+8)}{6} u_\perp^2 + \frac{n_\|}{6}u_\times^2 -
\frac{(3n_\perp + 14)}{12} u_\perp^3 - \frac{5n_\|}{36}u_\perp
u_\times^2
- \frac{n_\|}{9}u_\times^3,\\
\label{3} \beta_{u_\times} & = & -\varepsilon  u_\times +
\frac{(n_\perp+2)}{6}u_\perp u_\times + \frac{(n_\|+2)}{6}u_\times
u_\| + \frac{2}{3}u_\times^2 - \frac{(n_\perp+n_\|+16)}{72}
u_\times^3 \nonumber  \\
&-& \frac{(n_\perp+2)}{6}u_\times^2 u_\perp  - \frac{(n_\|+2)}{6}
u_\times^2  u_\| - \frac{5(n_\perp+2)}{72}
 u_\perp^2 u_\times  - \frac{5(n_ \|+2)}{72}u_\times u_\|^2, \\
\label{4} \beta_{u_\|} &=& -\varepsilon  u_\| + \frac{(n_\|+8)}{6}
u_\|^2 + \frac{n_\perp}{6}u_\times^2 - \frac{(3n_\| + 14)}{12}
u_\|^3 -\frac{5n_\perp}{36}u_\|u_\times^2 -
\frac{n_\perp}{9}u_\times^3.
\end{eqnarray}
Here, $\{u_\perp,u_\times,u_\|\}=\{u\}$ are renormalized couplings
and the space dimension $d$ enters the $\beta$-functions via
parameter $\varepsilon=4-d$. With the $\beta$-functions at hand, one
can analyze the flow equations of the fourth-order couplings
$\{u\}$:
\begin{equation}
\label{5} \ell\frac{du_a}{d\ell }=\beta_{u_a}(\{u\}) \, ,
\end{equation}
with $a=\perp,\, \|, \, \times$ and the flow parameter $\ell$, and
find the FPs $\{u^*\}$ of these equations as the solutions of the
system of equations
\begin{equation} \label{6}
\beta_{u_a}(\{u^*\})=0.
\end{equation}
Which of these FPs is the stable one depends on the number of
components $n_\perp$ and $n_\|$ and the dimension $d$ of space. The
scaling properties depend on the symmetry of stable FP.

\begin{figure}
\includegraphics[width=.7\textwidth]{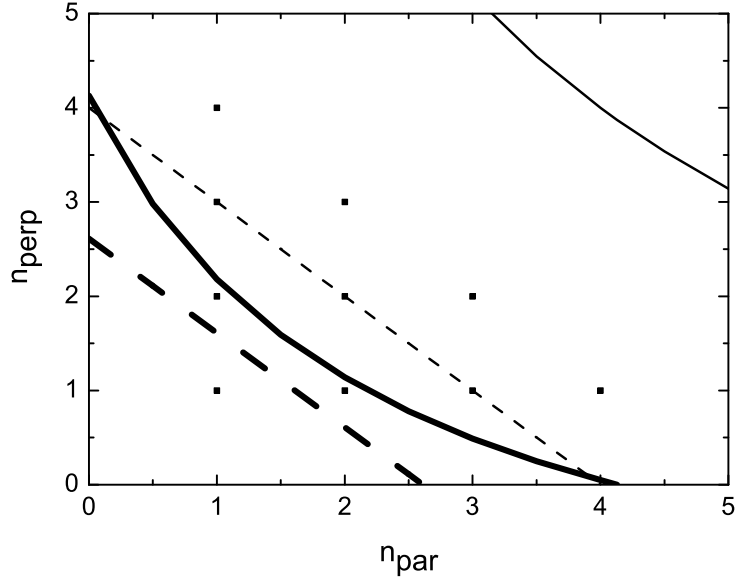}
 \caption{\label{fig2}  Regions of FPs stability in the
in the $n_\|-n_\perp$-plane, $d=3$. The lines separate regions where
Heisenberg FP ${\cal H}$, biconical FP ${\cal B}$ and decoupling FP
${\cal D}$ are stable (from left to right).
 Shown are the ${\cal HB}$-stability borderlines
(dashed lines) and ${\cal BD}$-stability borderlines (solid lines),
in   one loop order (thin lines) and resummed two loop order (thick
lines). The dots indicate low integer values for order parameter
components.}
\end{figure}

There are two alternative ways to look for the solutions of the FP
equations (\ref{6}) and, subsequently, for the scaling properties of
the system. In one approach, the $\varepsilon$-expansion, the
solutions are obtained as series in $\varepsilon$ and then evaluated
at the value of interest (at $\varepsilon=1$ for $d=3$ theories).
Alternatively, one may solve a system of non-linear equations
directly at the dimensionality of space of interest (e.g. at
$\varepsilon=1$) \cite{Schloms} and obtain the FP coordinates
numerically. The RG expansions being divergent \cite{rgbooks}, the
special resummation techniques are used to get convergent results
\cite{Holovatch02}. As we have discussed already above, depending on
the values of $n_\|,\, n_\perp,$ and $d$, the multicritical behavior
is governed by one of the three non-trivial FPs: ${\cal H}$
$\{u_\perp^*=u_\times^*=u_\|^*\}$, ${\cal B}$ $\{u_\perp^*\neq
u_\times^*\neq u_\|^*\}$, and ${\cal D}$ $\{u_\perp^*\neq 0,
u_\times^*=0, u_\|^*\neq 0\}$. In Fig. \ref{fig2} we show how the
stability of these FPs change with $n_\|,\, n_\perp,$   for $d=3$.
There, we compare the first order $\varepsilon$-expansion results
\cite{Lyuksyutov75,Kosterlitz76} with the two-loop results
\cite{Folk08a} obtained within the fixed $d=3$ technique
\cite{Schloms}. The two-loop results were obtained applying
Pad\'-Borel resummation technique to functions (\ref2)--(\ref{4})
\cite{note2}. One sees that the borderlines of the FPs stability are
drastically shifted to smaller values of OP components. Thus in the
case $n_|=1$ and $n_\perp=2$ FP ${\cal B}$ (connected with
tetracriticality) is stable in two loop order contrary to the one
loop calculations where the FP ${\cal H}$ (connected with
bicriticality) is stable. The resummed higher orders of the
perturbation theory do not change this result and do not lead to
essential changes in the critical exponents either
\cite{Calabrese03}.

As usually, the asymptotic values of the critical exponents are
defined by the stable FP values of the corresponding RG
$\zeta$-functions, which we do not expose here. Note that in
general, there are distinct exponents $\eta_\|$, $\eta_\perp$
governing spacial decay of the order parameter correlations in
directions parallel and perpendicular to the anisotropy axis. As a
consequence, there is a pair of $\gamma$-exponents, $\gamma_\|$,
$\gamma_\perp$ that govern corresponding isothermal magnetic
susceptibilities. However, the above RG procedure assumes that the
multicritical system is described by a single diverging length scale
and therefore by one correlation length $\xi$ and one corresponding
critical exponent $\nu$. This does not hold for decoupled systems
where two length scales  are present and the usual scaling laws with
one length scale break down \cite{Kosterlitz76}. We give typical
numerical values of the  exponents in Table \ref{tab1}.

\begin{table}
\centering \tabcolsep=2mm
\begin{tabular}{lllllll}
\hline
 Reference &
   FP & $\eta_\perp$ & $\eta_\|$ & $\gamma_\perp$ & $\gamma_\|$ & $\nu$  \\
   \hline
 \cite{Folk08a} &   ${\cal B}$ & 0.037 & 0.037 & 1.366 & 1.366 & 0.696         \\
 \cite{Folk08a} &   ${\cal H}(3)$ & 0.040 & 0.040 & 1.411 & 1.411 & 0.720               \\
 \cite{Nelson74} &  ${\cal B}$ & 0 & 0 & 1.222 & 1.222 & 0.611         \\
\cite{Nelson74} &   ${\cal H}(3)$ & 0 & 0 & 1.227 & 1.227 & 0.611       \\
\cite{Calabrese03} &   ${\cal B}$ & 0.037(5) & 0.037(5) & {\em
1.37(7)} & {\em 1.37(7)} & 0.70(3) \\
\cite{Guida98} &   ${\cal H}(3)$ & 0.0375(45) & 0.0375(45) &
1.382(9) & 1.382(9)  & 0.7045(55) \\
    \hline
\end{tabular}
\caption{Critical exponents of the $O(1)\oplus O(2)$ model obtained
in different approximations. \cite{Folk08a}: resummation of the
two-loop RG series at fixed $d=3$; \cite{Nelson74}: first order
$\varepsilon$-expansion, \cite{Calabrese03,Guida98}: resummed fifth
order $\varepsilon$-expansion. Numbers, shown in italic were
obtained via familiar scaling relations.
 \label{tab1}}
\end{table}

\begin{figure}[h]
\centerline{\includegraphics[width=.7\textwidth]{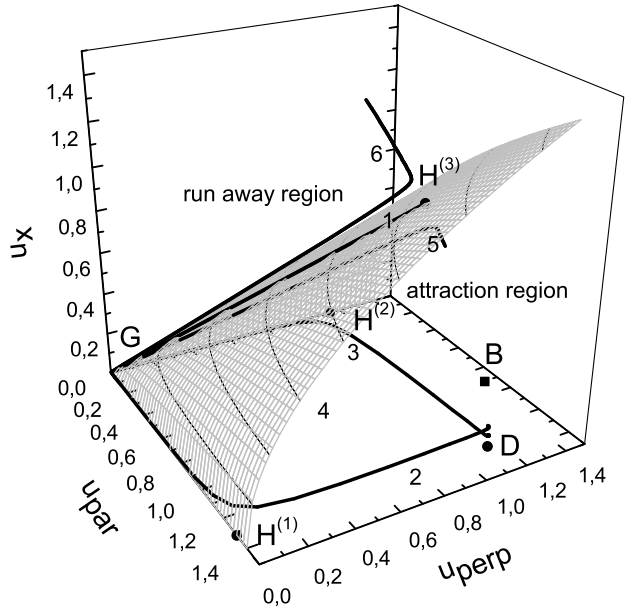}}
 \caption{\label{fig3}  Resummed RG flow of Eqs. (\ref{5})
     for different initial conditions at $d=3$, $n_\|=1$, $n_\perp=2$.
     The unstable FPs are shown as filled spheres, the
stable biconical FP as filled cube.  The FPs points are connected by
separatrices defining the surface which encloses the attraction
region.}
\end{figure}

Whereas the asymptotic critical exponent values are determined {\em
strictly at} the FP and correspond to the scaling behavior at the
multicritical point, of special interest are the effective critical
exponents which are observed {\em in the vicinity} of the
multicritical point. These are the effective exponents that often
are observed experimentally and are measured in MC simulations. In
the RG framework, one may estimate the effective exponents from the
values of corresponding RG $\zeta$-functions calculated along the RG
flow and relate the flow parameter $\ell$ to the distance to the
multicritical point. In Fig. \ref{fig3} we show the resummed
\cite{note2} RG flow of Eqs. (\ref{5}) for different initial
conditions \cite{Folk08a}. The unstable FPs are shown as filled
spheres, the stable biconical FP as filled cube. Let us note that
the neighborhood of the stability border lines to the other FPs
leads to very small transient exponents. Therefore, the stable FP is
not reached for the value of the flow parameter chosen in Fig.
\ref{fig3} (there, the flow parameter has been changed in the
interval $-40 \leq \ln \ell \leq 0$).

\begin{figure}
\tabcolsep=.01\textwidth
\begin{tabular}{cc}
\includegraphics[width=.5\textwidth]{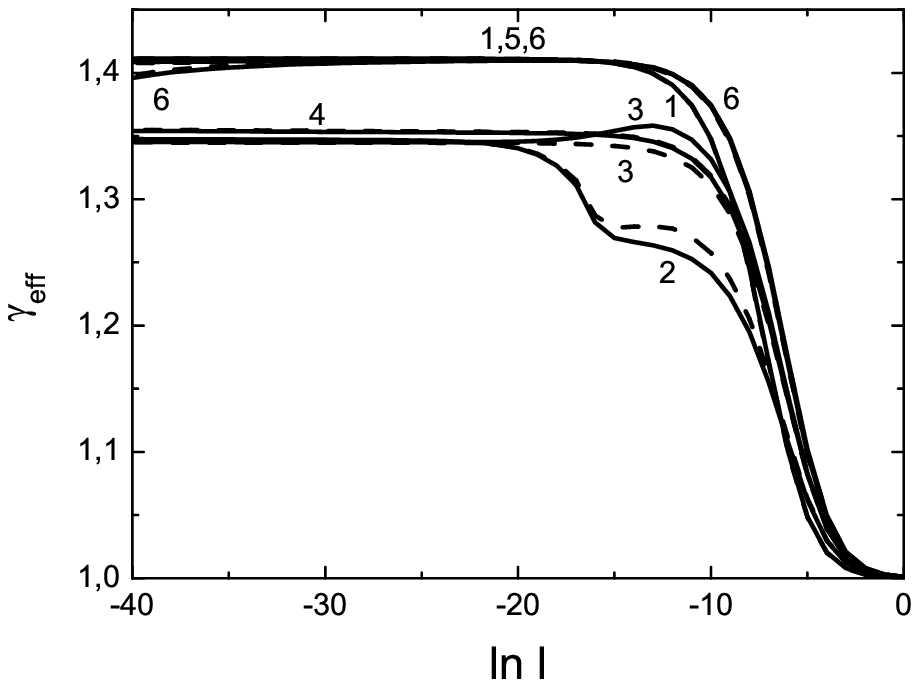} &
\includegraphics[width=.5\textwidth]{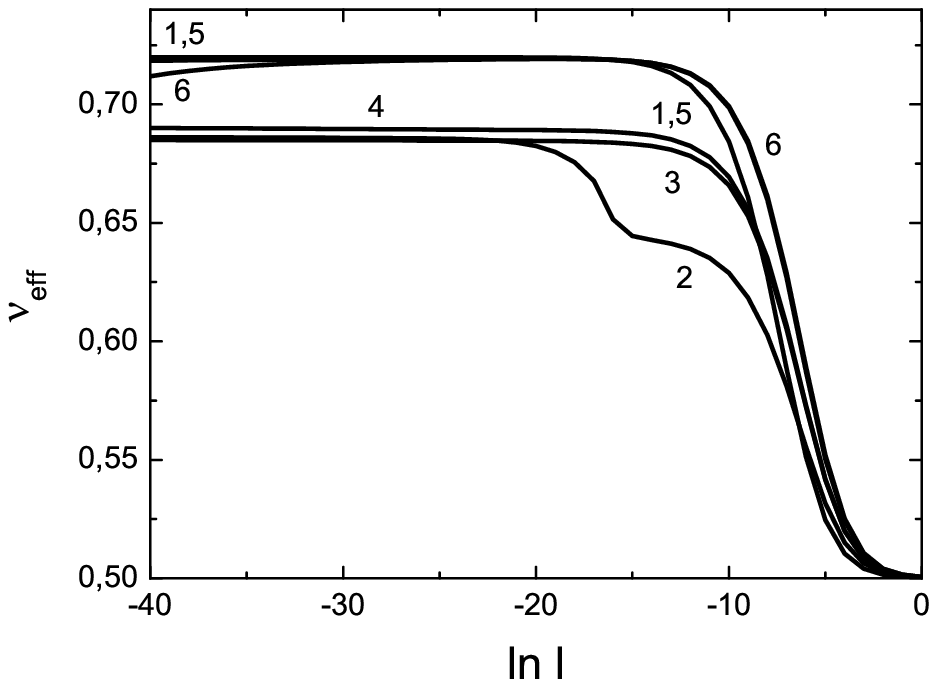} \\
{\bf a.} &  {\bf b.} \\
\end{tabular}
 \caption{\label{fig4} Effective exponents of different observables in the vicinity
 of a multicritical point for the the flows of Fig. \ref{fig3}. {\bf a}: isothermal
 magnetic susceptibility (solid curves: $\gamma_\|$, dashed curves: $\gamma_\perp$);
 {\bf b}: correlation length.}
\end{figure}

Defining the effective exponents as explained above, one can
evaluate their numerical values along the RG flows of Fig.
\ref{fig3} and in this way predict possible outcome of measuring the
scaling properties of different observables at multicritical point.
As two typical examples, we show in Fig. \ref{fig4} the change of
the values of isothermal susceptibility effective exponents
$\gamma_\|$, $\gamma_\perp$ and of the correlation length critical
exponent $\nu$ as the multicritical point is being approached, the
limit $T\to T_c$ corresponds to the limit $\ell\to\ 0$.

Before passing to discussion of peculiarities of dynamic
multicritical behavior,  let us note a particular feature of the
$O(n_\|)\oplus O(n_\perp)$ model that becomes evident from the above
analysis of the statics. As the stability analysis shows, for the
physically interesting case $d=3$, $n_\|=1$, $n_\perp=2$ the
asymptotic behavior is governed by the biconical FP ${\cal B}$.
Therefore, the tetracritical point is realized (c.f. Fig.
\ref{fig1}{\bf a}). However, depending on the particular microscopic
non-universal characteristics of a given system, one may expect a
variety of different scenarios for multicritical behavior, including
the triple point (that corresponds to the run away solutions of the
RG flow equations, c.f. Fig. \ref{fig3}) and bicritical point (when
for certain initial condition the Heisenberg FP ${\cal H}$ is
reached).

\section{Dynamics in the vicinity of multicritical points}

Sketched above particular features of static multicritical behavior
are further manifested  if the critical dynamics is addressed.
Below, we briefly analyze three different from of dynamical behavior
in the vicinity of multicritical points.

\subsection{Relaxational dynamics (model A)}

Let us start from the simplest dynamical model, model A, when one
assumes relaxational behavior for the two order parameters
$\vec{\phi}_\|$ and $\vec{\phi}_\perp$.  This model has been studied
in the one-loop approximation in \cite{Dohm77}, the two-loop results
have been obtained in \cite{Folk08a}. The model A type Langevin
equations of motion describe two order parameters that relax to
equilibrium with the relaxation rates (kinetic coefficients)
$\mathring{\Gamma}_\perp$ and $\mathring{\Gamma}_\|$:
\begin{eqnarray}
\label{7} \frac{\partial \vec{\phi}_{\perp 0}}{\partial
t}&=&-\mathring{\Gamma}_\perp
\frac{\delta {\mathcal H}}{\delta \vec{\phi}_{\perp 0}}+\vec{\theta}_{\phi_\perp} \ , \\
\label{8} \frac{\partial \vec{\phi}_{\|0}}{\partial
t}&=&-\mathring{\Gamma}_\| \frac{\delta {\mathcal H}}{\delta
\vec{\phi}_{\|0}}+\vec{\theta}_{\phi_\|} \, .
\end{eqnarray}
Here, ${\mathcal H}$ is the static effective Hamiltonian (\ref{1}),
index $0$ refers to bare (unrenormalized) quantities and the
stochastic forces $\vec{\theta}_{\phi_\perp}$,
$\vec{\theta}_{\phi_\|}$ fulfill Einstein relations
\begin{eqnarray}
\label{9} \langle\theta_{\phi_\perp}^\alpha(x,t)\
\theta_{\phi_\perp}^\beta (x^\prime,t^\prime)\rangle \!\!\!&=&\!\!\!
2\mathring{\Gamma}_\perp\delta(x-x^\prime)\delta(t-t^\prime)\delta^{\alpha\beta}
\ , \\
\label{10} \langle\theta_{\phi_\|}^i(x,t)\
\theta_{\phi_\|}^j(x^\prime,t^\prime) \rangle \!\!\!&=&\!\!\!
2\mathring{\Gamma}_\|\delta(x-x^\prime)\delta(t-t^\prime)
\delta^{ij} \ ,
\end{eqnarray}
with indices $\alpha,\beta=1,\dots , n_\perp$ and $i,j=1,\dots
,n_\|$ corresponding to the two subspaces.

Application of the RG procedure to study dynamical multicritical
behavior relies on the Bausch-Janssen-Wagner approach
\cite{Bausch76}, where the appropriate Lagrangian of the model is
studied and dynamic vertex functions are calculated in perturbation
theory and renormalized. In such a technique, essential
simplification of calculations is achieved due to the possibility to
single out a static part of every dynamic vertex function
\cite{Folk02,Folk06}. Renormalization of the kinetic coefficients
gives rise to appropriate $\beta$-functions. Here, we reveal the
two-loop $\beta$-function for the time-scale ratio
$v=\Gamma_\|/\Gamma_\perp$ between the renormalized kinetic
coefficients  $\Gamma_\|$ and $\Gamma_\perp$. The function reads
\cite{Folk08b}:
\begin{eqnarray}
\beta_v&=&\frac{v}{72}\Bigg\{\Big[(n_\|+2)u_\|^2-(n_\perp+2)u_\perp^2\Big](6\ln\frac{4}{3}-1)
\nonumber -n_\|\ u_\times^2\left[\frac{4}{v}\ln\frac{2(1+v)}{2+v}
+2\ln\frac{(1+v)^2}{v(2+v)}-1\right]   \\ \label{11} &+&n_\perp\
u_\times^2\left[4v\ln\frac{2(1+v)}{1+2v}
+2\ln\frac{(1+v)^2}{1+2v}-1\right]\Bigg\} \, .
\end{eqnarray}
As we have noted in the preceding section discussing the static
critical behavior, a non universal effective critical behavior may
be observed if the values of the static couplings and the time scale
ratio are not in a FP but rather are described by the flow
equations. For $v$ the flow equation reads
\begin{equation}  \label{12}
\ell \frac{d v}{d
\ell}=\beta_v\big(u_\|(\ell),u_\perp(\ell),u_\times(\ell),v(\ell)\big)
\, .
\end{equation}
Below we will show some results about non-universal dynamic
multicritical behavior obtained with two-loop accuracy. The
numerical results for the static part of the RG function were
obtained by means of the resummation technique \cite{note2}, whereas
no resummation has been applied to the dynamic functions
\cite{Folk08b}.

\begin{figure}
\includegraphics[width=.7\textwidth]{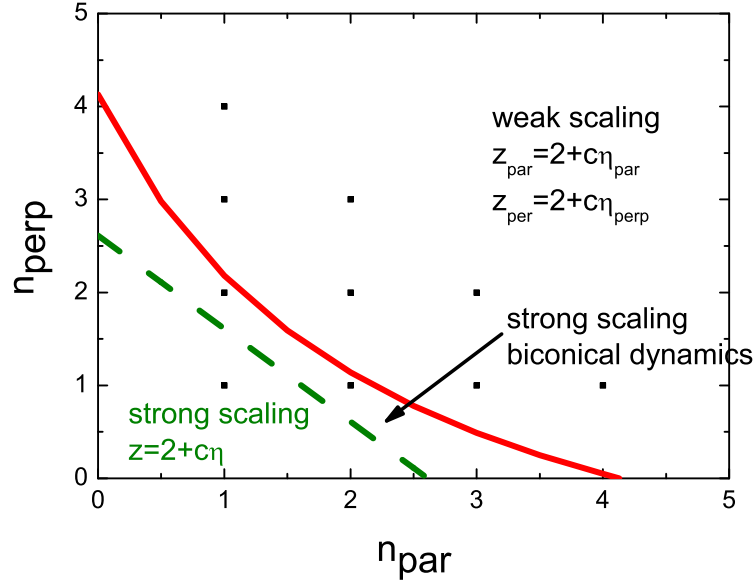}
 \caption{\label{fig5}  Regions of the different types of dynamic
 scaling behavior, $\epsilon=4-d=1$. The rest of notations are as
 in  Fig. \ref{fig2}.}
\end{figure}

\begin{figure}
\tabcolsep=.01\textwidth
\begin{tabular}{cc}
\includegraphics[width=.5\textwidth]{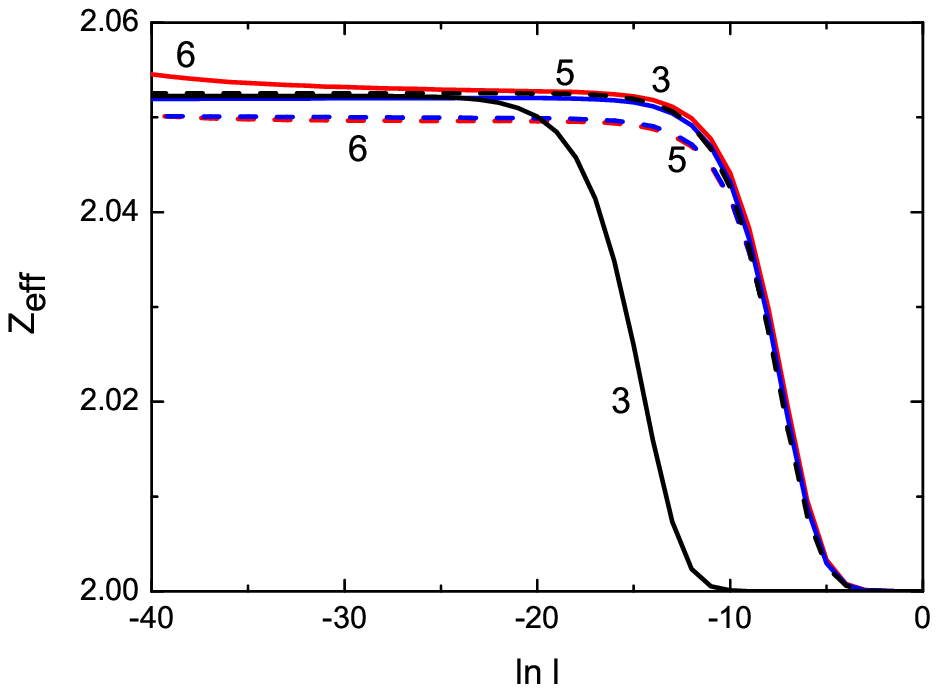} &
\includegraphics[width=.5\textwidth]{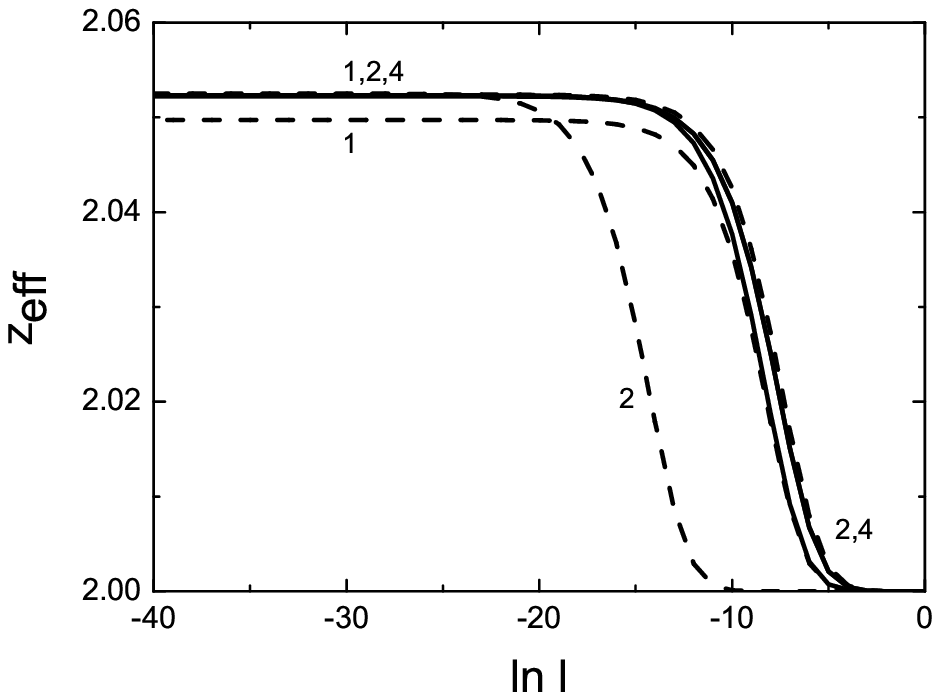} \\
{\bf a.} &  {\bf b.} \\
\end{tabular}
 \caption{\label{fig6} Model A multicritical dynamics. Effective dynamical exponent in for different RG flows
in the vicinity of a multicritical point at $d=3$, $n_\|=1$,
$n_\perp=2$. The labeling of the flows corresponds to Fig.
\ref{fig3}. The exponents for the perpendicular (dashed curves) and parallel (solid curves)
components of the OP differ in the non asymptotic region.}
\end{figure}

One of the quantities of interest that characterize dynamic critical
phenomena is the autocorrelation time $\tau$. It is known to diverge
as the critical point $T_c$ is approached, the divergency is
described by the power law:
\begin{equation}  \label{13}
\tau \sim |T-Tc|^{-\nu z},
\end{equation}
with the universal correlation length and dynamic critical exponents
$\nu$ and $z$, correspondingly. In the multicritical phenomena we
consider, one distinguishes two dynamical critical exponents, $z_\|$
and $z_\perp$, that govern the power law increase of the
autocorrelation time for the order parameters $\vec{\phi}_\|$ and
$\vec{\phi}_\perp$, correspondingly. In asymptotics they  are
defined by the stable FP values of the corresponding RG functions.
At the strong scaling FP there is only one dynamic time scale and
the two exponents are equal whereas at the weak scaling FP they are
different and define for each component, parallel and perpendicular,
the time scale. As it follows from our calculations \cite{Folk08b}
and as one may see from the Fig. \ref{fig5}, the region of stability
of the biconical FP ${\cal B}$ (physically important case $d=3$,
$n_\|=1$, $n_\perp=2$ belongs to this region) is characterized by
the strong scaling dynamics: the time relaxation of both order
parameters, $\vec{\phi}_\|$ and $\vec{\phi}_\perp$  is governed by
the same exponent. In Fig. \ref{fig6} we show an evolution of this
exponent $z_{\rm eff}$ to its asymptotic value $z=2.05$ when the
time-scale ratio $v$ is set to its FP value and the static couplings
$u$ change along the RG flows of Fig. \ref{fig3}. Since the exponents have not
reached their (equal) asymptotic values differences between the parallel and
perpendicular components of the OP remain.

\subsection{Conservation of magnetization (model C)}

A step towards making the description of dynamic phenomena in the
vicinity of a multicritical point more realistic is to take into
account possible couplings between the order parameters and
conserved densities, that is to consider the model C dynamics
\cite{HHreview,Halperin74}. In the problem under consideration,
there are two types of conserved densities: one is
magnetization-like (more precisely, it is the parallel component of
the magnetization), another is the energy density. We will not
consider this second density here, as far as up to the two-loop
order the specific heat critical exponent $\alpha$ is negative for
the case $d=3$, $n_\|=1$, $n_\perp=2$ which is of most interest
here. Therefore, a coupling to the energy density is irrelevant in
the RG sense - it vanishes at the FP \cite{Folk06}. An account of
both the order parameter and the (conserved) scalar density  is
achieved by an extension of the static functional (\ref{1}). Now,
the corresponding model C static functional reads:
\begin{equation}\label{14}
{\cal H}^{(C)}\!=\! {\cal H} + \int\! d^dx\Bigg (\frac{1}{2}m_0^2
+\frac{1}{2}\mathring{\gamma_\perp}m_0\vec{\phi}_{\perp
0}\cdot\vec{\phi}_{\perp 0}
+\frac{1}{2}\mathring{\gamma_\|}m_0\vec{\phi}_{\|
0}\cdot\vec{\phi}_{\| 0} -\mathring{h}m_0\Bigg ) \ .
\end{equation}
Here, the first term in the right hand side is given by Eq.
(\ref{1}), the density $m_0\equiv m_0(x)$ is a scalar quantity,
$\mathring{h}$ is a field conjugated to $m_0$,
$\mathring{\gamma_\perp}$ and $\mathring{\gamma_\|}$  are asymmetric
static couplings  between the corresponding order parameters and the
conserved density.

In their turn, the relaxational equations of motion
(\ref{7}),(\ref{8}) are now extended by including a diffusion
equation for the scalar density:
\begin{eqnarray}
\label{15} \frac{\partial \vec{\phi}_{\perp 0}}{\partial
t}&=&-\mathring{\Gamma}_\perp \frac{\delta {\mathcal
H}^{(C)}}{\delta \vec{\phi}_{\perp 0}}+\vec{\theta}_{\phi_\perp}
\ , \\
\label{16} \frac{\partial \vec{\phi}_{\|0}}{\partial
t}&=&-\mathring{\Gamma}_\| \frac{\delta {\mathcal H}^{(C)}}{\delta
\vec{\phi}_{\|0}}+\vec{\theta}_{\phi_\|}
\, , \\
\label{17} \frac{\partial m_0}{\partial
t}&=&\mathring{\lambda}\nabla^2 \frac{\delta {\mathcal
H}^{(C)}}{\delta m_0}+\theta_m \, .
\end{eqnarray}
Here, the static functional ${\mathcal H}^{(C)}$ is given by
(\ref{14}), $\mathring{\lambda}$ is a kinetic coefficient of
diffusive type for the scalar density, the rest of notations is as
in (\ref{7}),(\ref{8}). The stochastic forces
$\vec{\theta}_{\phi_\perp}$, $\vec{\theta}_{\phi_\|}$ satisfy the
Einstein relations (\ref{9}), (\ref{10}), with an additional
Einstein relation for the new stochastic force  $\theta_m$:
\begin{equation}\label{18}
\langle\theta_m(x,t)\ \theta_m(x^\prime,t^\prime) \rangle
\!\!\!=\!\!\!
-2\mathring{\lambda}\nabla^2\delta(x-x^\prime)\delta(t-t^\prime) \,
.
\end{equation}
The renormalization of the above introduced asymmetric couplings
$\mathring{\gamma_\perp}$, $\mathring{\gamma_\|}$ and kinetic
coefficient $\mathring{\lambda}$ leads to new RG functions. In
particular the RG flow of the time scale ratios
\begin{equation}\label{19}
w_\perp=\frac{\Gamma_\perp}{\lambda} \ , \qquad
w_\|=\frac{\Gamma_\|}{\lambda}
\end{equation}
is now governed by the appropriate functions $\beta_{w_\perp}$ and
$\beta_{w_\|}$, correspondingly. Note that defined for model A time
scale ratio $v$ is equally well defined in terms of (\ref{19}):
\begin{equation}\label{20}
v\equiv\frac{\Gamma_\|}{\Gamma_\perp}=\frac{w_\|}{w_\perp} \, .
\end{equation}
Therefore, the dynamical FP equations:
\begin{equation}\label{21}
\beta_{w_\perp}(w^*_\perp,w^*_\|,v^*)=
\beta_{w_\|}(w^*_\perp,w^*_\|,v^*)=
\beta_{v}(w^*_\perp,w^*_\|,v^*)=0
\end{equation}
are now not independent: one of these equations can be eliminated by
the relation (\ref{20}).

\begin{figure}
\tabcolsep=.01\textwidth
\begin{tabular}{cc}
\includegraphics[width=.4\textwidth]{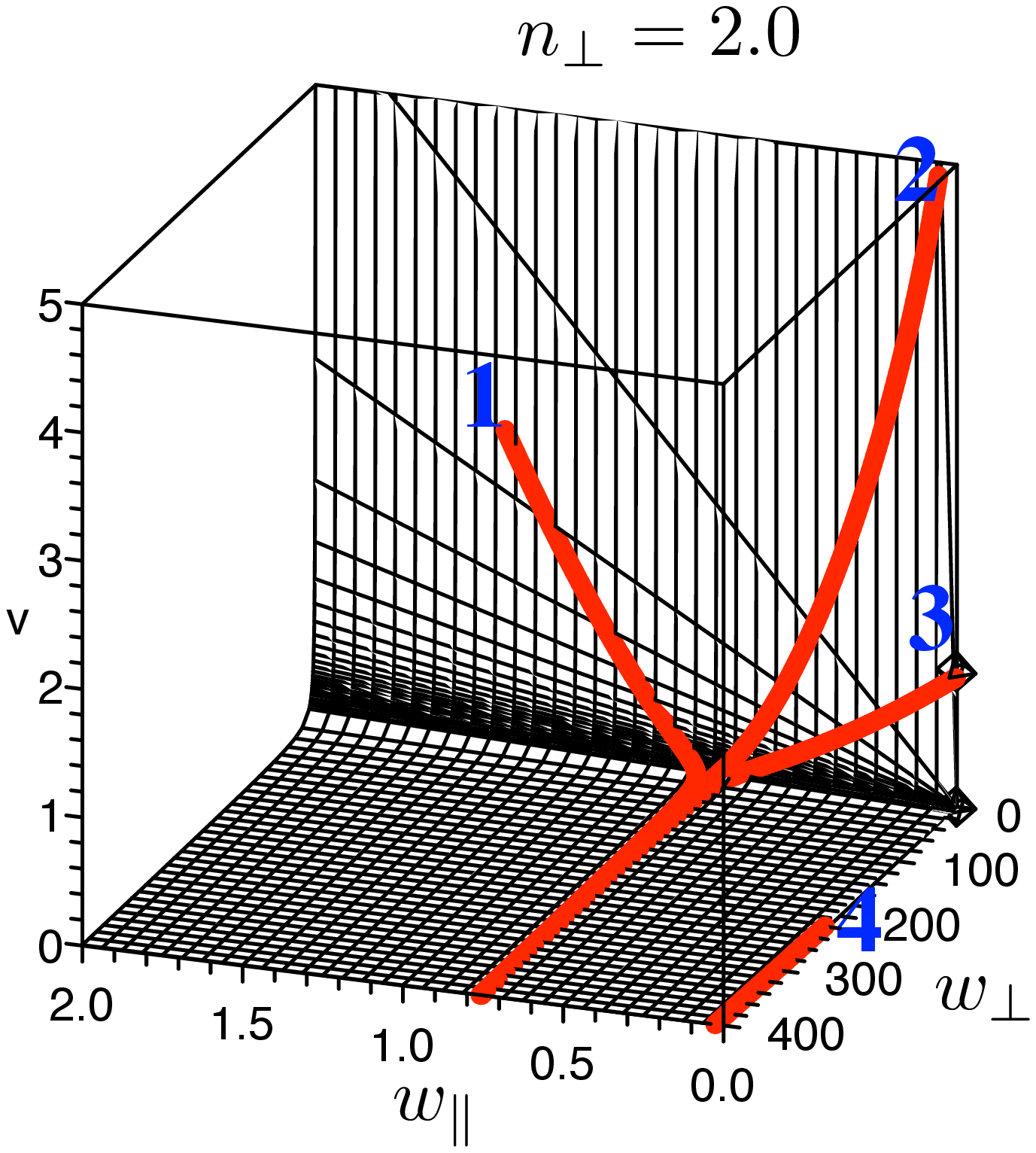} &
\includegraphics[width=.5\textwidth]{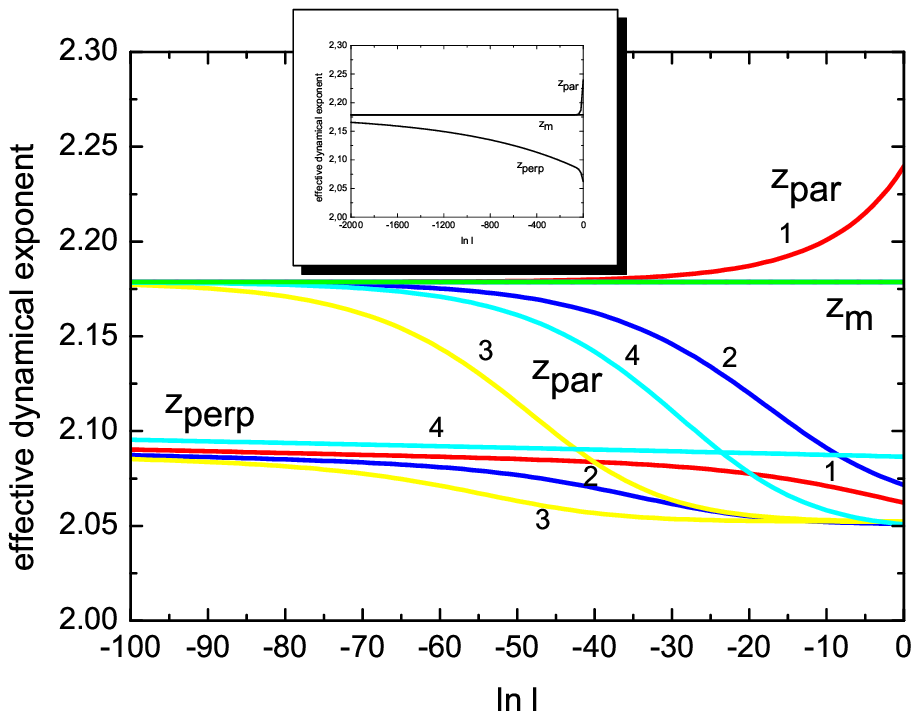} \\
{\bf a.} &  {\bf b.} \\
\end{tabular}
 \caption{\label{fig7} Model C effective dynamical multicritical behavior at $d=3$, $n_\|=1$,
$n_\perp=2$. {\bf a}: dynamical RG flow for different initial
conditions numbered from 1 to 4. {\bf b}: effective dynamical
exponents $z_\|$, $z_\perp$, and $z_m$ calculated along the RG flows
of Fig. \ref{fig7}{\bf a}, as indicated by the numbers.}
\end{figure}

Equations of motion (\ref{15})--(\ref{17}) describe time evolution
of three different observables. Each of them has its own
autocorrelation time  which, as the multicritical point is reached,
may be governed by an independent dynamical critical exponent. In
addition to the two exponents defined in the former subsection,
$z_\|$ and $z_\perp$, the dynamical critical exponent $z_m$ for the
scalar density is to be considered. Similar, as in the model A case,
these three exponents may coincide, in the strong scaling dynamical
FP or they may differ, in the weak  scaling dynamical FP. Complete
stability analysis of the model C RG equations in two-loop
approximation is given in Ref. \cite{Folk08b}. In particular, it is
shown that for the case  $d=3$, $n_\|=1$, $n_\perp=2$, where the
static FP is the biconical FP ${\cal B}$, the strong scaling
dynamical FP is stable. Physically this means that in asymptotics
the multicritical dynamics is characterized by one time scale, and
three dynamical exponents coincide. In particular, their
asymptotical value was found to be $z_\|=z_\perp=z_m=2.18$
\cite{Folk08a}. However, as it was revealed in the former sections,
the effective multicritical behavior is much reacher. In particular,
in Fig. \ref{fig7}{\bf a} we show the RG flows calculated for
different dynamical initial conditions when the static couplings are
chosen to be fixed at their biconical FP values. The stable
dynamical (strong scaling) FP lies outside the region shown. Also
shown is the surface $v=w_\|/w_\perp$ to which the flow is
restricted by the condition (\ref{20}). The RG flows of Fig.
\ref{fig7}{\bf a} give rise to difference in the effective dynamical
critical exponents, as shown in Fig. \ref{fig7}{\bf b}. The insert
of the figure shows that even for flow parameters as small as $\ln
\ell=-2000$ the effective exponent $z_\perp$ has not reached its
asymptotic value $2.18$.

\subsection{The complete dynamic model (model G)}

We now restrict ourselves to the case of $n_\|=1$, $n_\perp=2$ and
include mode coupling terms, which correspond to Larmor terms
describing the precession of the alternating magnetization  and the
magnetization around each other. They are well known from the
isotropic antiferromagnet without an external field \cite{mazenko}.
Then within an external magnetic field the corresponding equations
read
\begin{eqnarray}
\label{modG}
\frac{\partial \phi_{\perp 0}^\alpha}{\partial t}&=&-\mathring{\Gamma}^\prime_\perp
\frac{\delta {\mathcal H}^{(C)}}{\delta \phi_{\perp 0}^\beta}
+\mathring{\Gamma}^{\prime\prime}_\perp \epsilon^{\alpha\beta z}
\frac{\delta {\mathcal H}^{(C)}}{\delta \phi_{\perp 0}^\alpha}
+\mathring{g}\ \epsilon^{\alpha\beta z}
\phi_{\perp 0}^\beta\frac{\delta {\mathcal H}^{(C)}} {\delta m_0}+\theta_{\phi_\perp}^\alpha
\ , \\
\frac{\partial \phi_{\|0}}{\partial t}&=&-\mathring{\Gamma}_\|
\frac{\delta {\mathcal H}^{(C)}}{\delta \phi_{\|0}}+\theta_{\phi_\|}
\, , \\
\frac{\partial m_0}{\partial t}&=&\mathring{\lambda}\nabla^2
\frac{\delta {\mathcal H}^{(C)}}{\delta m_0}+
\mathring{g}\ \epsilon^{z\alpha\beta}\phi_{\perp 0}^\alpha
\frac{\delta {\mathcal H}^{(C)}}{\delta\phi_{\perp 0}^\beta}
 +\theta_m \, .
\end{eqnarray}
Now $\alpha$ and $\beta$ indicates the planar components $x,y$ and the Levi-Civita tensor
$\varepsilon^{\alpha \beta z}$ with the third index fixed to $z$ has been introduced. The
parallel component of the OP is its $z$-component. This component remains just relaxing,
whereas the planar components of the OP are coupled to the $z$-component of the magnetization
by the precession terms.

A new feature arises because of the simultaneous presence of the
mode coupling $\mathring{g}$ and the asymmetric static couplings
$\mathring{\gamma_\perp}$ and $\mathring{\gamma_\|}$ in ${\cal
H}^{(C)}$ (\ref{14}). The perpendicular relaxation coefficient
$\mathring{\Gamma}_\perp$ has to be considered a complex quantity
where the imaginary part constitute a precession term (second term
on the right hand side of (\ref{modG}). Even if in the background
such terms are absent they are produced by the renormalization
procedure.

The stochastic forces $\vec{\theta}_{\phi_\perp}$, $\theta_{\phi_\|}$
and $\theta_m$ fulfill Einstein relations
\begin{eqnarray} \label{thetaperp}
\langle\theta_{\phi_\perp}^\alpha(x,t)\ \theta_{\phi_\perp}^\beta
(x^\prime,t^\prime)\rangle \!\!\!&=&\!\!\!
2\mathring{\Gamma}^\prime_\perp\delta(x-x^\prime)\delta(t-t^\prime)\delta^{\alpha\beta}
\ , \\
\label{thetapara}
\langle\theta_{\phi_\|}(x,t)\ \theta_{\phi_\|}(x^\prime,t^\prime)
\rangle \!\!\!&=&\!\!\! 2\mathring{\Gamma}_\|\delta(x-x^\prime)\delta(t-t^\prime)
\ , \\
\label{thetam}
\langle\theta_m(x,t)\ \theta_m(x^\prime,t^\prime)
\rangle \!\!\!&=&\!\!\! -2\mathring{\lambda}\nabla^2\delta(x-x^\prime)\delta(t-t^\prime)
\ .
\end{eqnarray}

This model has been solved in one loop order in \cite{Dohm77} using
the one loop results of statics. As is has been already seen for the
simpler dynamic models models changes are expected in two loop order
both by the statics as well as by the dynamic terms especially of
model C type. We have calculated the complete field theoretic
functions in two loop order \cite{Folk09b} necessary to calculate
the critical (effective) dynamical exponents. Independent whether
the Heisenberg or biconical  is the stable static FP a first
inspection of the flow of the dynamical parameters shows the
following: (i) The imaginary part of the perpendicular relaxation
rate renormalizes to zero, (ii) the times scale ratios $v$
(\ref{20}),  $w_\|$ (\ref{19}) approach zero and $w_\perp$ increases
to $\infty$. Irrespective of the kind of the stable dynamic FP -
wether it is a strong scaling FP with very small but finite or a
weak scaling FP with zero values for $v$ and $w_\|$ - the physical
observable features of the magnetic transport coefficient are
effective ones. The range of effective values for the dynamic
exponents corresponding to the relaxation of the perpendicular and
parallel alternating magnetization and the magnetization are
starting around its Van Hove values $z_\perp\sim z_\|\sim
z_\lambda\sim 2$ in the background and approach for the biconical FP
deep in the asymptotic regime
\begin{equation}
 z_\perp \sim 1.6  \qquad z_\|\sim 2 \qquad z_\lambda\sim 1.6 \, .
\end{equation}
The main prediction according to this result would be that the perpendicular and the
parallel component of the OP would scale differently in this region.

The importance of this magnetic system  lies in the physical
accessibility of the OP, contrary to superfluid $^4$He or superfluid
mixture of $^4$He and $^3$He whose dynamics is described by model F
\cite{HHS76}. Here all quantities are in principle measurable
quantities.  Thus the prediction of the different dynamic scaling of
the OP components can be tested.

\section{Conclusions and outlook}

By this review we wanted to summarize recent progress achieved in
theoretical description of the multicritical phenomena. Whereas
traditionally RG techniques address critical points in their
different realizations, the description of multicritical phenomena
is possible both on quantitative and accurate qualitative levels.
Moreover, the problem appears to be tractable analytically even if
the complicated forms of multicritical dynamics are confronted. As
is revealed by the theoretical analysis, a particular feature of
static and dynamic behavior inherent to multicritical points is the
multitude of fixed points that describe the RG flow. In its turn,
this gives rise to rich effective behavior that may be characterized
by different types of multicritical points. A natural continuation
of performed studies would be to analyze cumulative effects caused
on the multicritical behavior by symmetry breaking factors of
different forms (single-ion anisotropies, disorder, frustrations)
that might be present in a system.

\begin{theacknowledgments}
We thank A. Fedorenko and W. Selke for useful discussions. Yu. H.
thanks B. Berche for his kind hospitality and an  inspiring
atmosphere at the Statistical Physics Group, Institut Jean Lamour,
University of Nancy, where this paper has been finalized. This work
was supported by the Fonds zur F\"orderung der wissenschaftlichen
Forschung under Project No. P19583-N20 and by the French-Ukrainian
bilateral collaboration Project "Dnipro".
\end{theacknowledgments}


\end{document}